\documentclass[12pt]{article}
\pdfoutput=1
\usepackage{epsfig}
\usepackage[utf8]{inputenc}
\usepackage{epsfig}
\usepackage{amsfonts}
\usepackage{amssymb}
\usepackage{mathtools}
\usepackage{amsmath}
\usepackage{cancel}
\usepackage{color}
\usepackage{slashed}
\usepackage{cite}
\usepackage{calligra}
\usepackage{caption}
\usepackage{subcaption}
\usepackage{comment}
\usepackage{marginnote}
\usepackage{accents}
\usepackage{url}

\allowdisplaybreaks

\begin{document}

\begin{center}

\vspace{1cm}

\renewcommand{\thefootnote}{\fnsymbol{footnote}}

{\bf \Large Infrared properties of five-point massive amplitudes in $\mathcal{N}=4$ SYM on the Coulomb branch} \vspace{1.0cm}

{\large L.V. Bork$^{1,2}$, N.B. Muzhichkov$^{2,3}$, E.S. Sozinov$^{2,4}$}
\footnote{E-mail:
~\rm{bork@itep.ru},~\rm{muzhnikita@gmail.com},~\rm{e.s.sozinov@gmail.com}}
\vspace{0.5cm}

{\it
$^1$Institute for Theoretical and Experimental Physics, Moscow, Russia,\\
$^2$The Center for Fundamental and Applied Research, \\ All-Russia
Research Institute of Automatics, Moscow, Russia, \\
$^3$Institute for Theoretical and Mathematical Physics, MSU, Moscow, Russia,\\
$^4$National Research Nuclear University MEPhI, Moscow, Russia
}

\renewcommand{\thefootnote}{\arabic{footnote}}
\setcounter{footnote}{0}

\vspace{1cm}

\abstract{We investigate the structure of the five-point $W$ boson scattering amplitude in $\mathcal{N}=4$ SYM on the Coulomb branch in a small mass limit. We show that up to two loops the IR divergences exponentiate and are controlled by the $\Gamma_{oct}$ anomalous dimension similar to the four-point amplitude case considered recently in the literature. We also make a conjecture regarding the all-loop structure of the five-point amplitude.}
\end{center}

\begin{center}
\end{center}

\newpage

\tableofcontents{}\vspace{0.5cm}

\section{Introduction}\label{s1}
In the last decades, there has been substantial progress in understanding the structure of amplitudes in supersymmetric gauge theories. The planar $\mathcal{N}=4$ SYM theory is the most prominent example. See for a review \cite{AldayAmplWloopRev,Henrietta_Amplitudes,Talesof1001Gluons} and references therein. Modern powerful analytical computational approaches \cite{Henrietta_Amplitudes} together with integrability methods for amplitudes \cite{OPE_for_W_Loops,Basso:2010in_GKP_string,Basso:2013_I,Basso:2013aha_II,Basso:2014koa_III,Basso:2014nra_IV,Basso:2014hfa_All_Helicity_I,Basso:2015rta_All_Helicity_II,Basso:2013vsa_MatrixPart} and correlation functions \cite{Hex1,Hex2,Basso:2017khq} allowed one to obtain a plethora of all-loop results or results at very high orders of perturbation theory that were nearly impossible to obtain using standard Feynman Diagram-based methods. 

Most of these results for amplitudes were related to the situation with unbroken gauge symmetry when all on-shell states (particles) from the $\mathcal{N}=4$ SYM supermultiplet are massless. However, one can use freedom in choosing Vacuum Expectation Values (VEVs) of scalar fields from the $\mathcal{N}=4$ SYM Lagrangian and consider $\mathcal{N}=4$ SYM with spontaneously broken gauge symmetry when some particles become massive. One can use this setup as an alternative way to regularise IR divergences in scattering amplitudes \cite{HennGiggs1,HennGiggs2,HennGiggs3} that manifest themselves as powers of $\log(m^2)$ in a small mass limit. There are also some hints that  integrability will survive in the massive case to some extent. 

The recent results of \cite{Caron-Huot:2021usw} suggest that there is likely a new duality between the amplitudes with massive external particles and the correlation functions of 1/2-BPS operators with  large R-charge. These correlation functions can (in principle) be evaluated to all loop orders using integrability methods \cite{Hex1,Hex2,Basso:2017khq} which allowed the authors of \cite{Caron-Huot:2021usw,Coronado:2018cxj} to obtain an all-loop result for the four-point correlation function which was conjectured to be dual to the four-point massive amplitude. But what was, probably, more important, the results of \cite{Caron-Huot:2021usw} revealed that the IR divergences in this massive case are \emph{not} controlled by the $\Gamma_{cusp}$ anomalous dimension which is in tension with what was expected previously \cite{ConformalProperties4point}. It turned out that the IR divergences in this case are controlled by a different known function $\Gamma_{oct}$ that emerged in the computation of the four-point large $R$-charge correlator. 

In the present article, we investigate the situation with the structure of IR divergences in the case of $n=5$ point massive $W$ boson amplitudes in planar $\mathcal{N}=4$ SYM with spontaneously broken gauge symmetry. We present indirect arguments regarding the structure of the master integral expansion of the five-point amplitude of massive particles in a small mass limit at one and two loops. We then evaluate these integrals in a single-scale kinematic limit and show that the IR divergences exponentiate and are also controlled by $\Gamma_{oct}$ (at least at this loop order). We then show that in combination with the assumption about the functional dependence on the kinematic invariants of the five-point amplitude the exponentiation of IR divergences allows one to determine the finite part of the amplitude up to a kinematic independent constant. This suggests the simple exponentiation pattern for the $n=4,5$ amplitudes which can be considered as another version of the BDS ansatz. 

This article is organised as follows. In Section \ref{s2}, we briefly discuss the structure of IR divergences of planar colour-ordered amplitudes in pure massless and massive cases in the $\mathcal{N}=4$ SYM theory.

In Section \ref{s3}, we discuss a possible structure of the integrands of the $n=5$ point amplitude and show that after the evaluation of the corresponding integrals the IR divergences exponentiate and this will likely imply the exponentiation of the finite part as well. The $n=4$ massive amplitude is also discussed for completeness as well as the two-point form factors of the operators from the stress-tensor supermultiplet.

In Conclusion we once again summarise the main results. In Appendices, we give details regarding loop integral computations and also explicitly show the
relations between $\Gamma_{oct}$ and $\Gamma_{cusp}$ anomalous dimensions. 

\section{IR properties of amplitudes on the Coulomb branch in $\mathcal{N}=4$ SYM}\label{s2}
\subsection{Massless amplitudes}\label{s21}
In addition to the coupling constant\footnote{We define the coupling constant as $a=g_{YM}^2N_c/(8\pi^2)$.} $a$ the $SU(N_c)$ planar $\mathcal{N}=4$ SYM  has other free adjustable parameters ("moduli") - the VEVs of the six real scalar fields $\phi^{AB}$ from the Lagrangian of the theory. These VEVs are usually considered equal to zero (i.e. we are considering the theory at the origin of its moduli space), but there are no restrictions on their values and one can consider a theory with non-zero VEVs. This case is often called in the literature the $\mathcal{N}=4$ SYM theory on the Coulomb branch \cite{HenrietteCoulomb,Henn:2010kb}.

As was pointed out in the introduction, in the last decades lots of results regarding the general structure of amplitudes including their IR behaviour in $\mathcal{N}=4$ SYM were obtained. Most of these results were dedicated to the amplitudes at the origin of the moduli space i.e. to the massless case. As in any gauge theory with massless particles the amplitudes in $\mathcal{N}=4$ SYM at the origin of the moduli space possess IR divergences. One can regularize these 
divergences via dimensional regularisation considering the theory in $D=4-2\epsilon$. The divergences manifest themselves as $1/\epsilon$ poles. The structure of these divergences is rather well-understood and one can formulate the following statement \cite{Mueller:1979ih,Magnea:1990zb,Sterman:2002qn,BernBDS}.
One can define the so called MHV amplitudes based on the total helicity of the particles participating in the scattering. Then for the colour-ordered MHV amplitudes $A_n$ one can show that to all loop orders the following relations hold \cite{BernBDS}:
\begin{eqnarray}\label{LogMDimReg}
\frac{A_n}{A_n^{tree}}&=&M_n=\sum_{l=0}^{\infty}a^lM_{n}^{(l)},\nonumber\\
\log M_n &=& -\sum_{l=1}^{\infty}\sum_{i=1}^n\frac{a^l}{4}\left(\frac{\Gamma^{(l)}_{cusp}}{(l\epsilon)^2}+\frac{G^{(l)}}{(l\epsilon)}\right)\left(\frac{\mu^2}{s_{ii+1}}\right)^{l\epsilon}
+\mathcal{F}_n+O(\epsilon),
\end{eqnarray}
where the coefficients $\Gamma^{(l)}_{cusp}$ and $G^{(l)}$ define two functions, the Cusp anomalous dimension $\Gamma_{cusp}(a)$ and the Collinear anomalous dimension $G(a)$, $s_{ii+1}$ are the squares of the sum of the consecutive momenta $s_{ii+1}=(p_i+p_{i+1})^2$ and $\mu$ is the mass parameter of the dimensional regularisation;
$\mathcal{F}_n$ is the function of the coupling constant and kinematical invariants only. It represents the finite, IR regulator independent, part of the amplitude. The $1/\epsilon^2$ pole structure originates from the combination of soft and collinear divergences, where each of them manifests itself as $1/\epsilon$.

From this relation one can see that IR divergences in all MHV amplitudes exponentiate\footnote{IR divergences in all other amplitudes in $\mathcal{N}=4$ SYM also exponentiate in a similar way.} with "critical exponents" given by $\Gamma_{cusp}(a)$ and $G(a)$. The coefficients $G^{(l)}$ are scheme-dependent, while $\Gamma^{(l)}_{cusp}$ are not. The first orders of perturbative expansion of $\Gamma_{cusp}(a)$ and $G(a)$ are given by: 
\begin{eqnarray}\label{GCuspAndGPT}
\Gamma_{cusp}(a)&=&\sum_{l=1}^{\infty}\Gamma_{cusp}^{(l)}a^l=2a-2\zeta_2a^2+11\zeta_4a^3+\ldots,\nonumber\\
G(a)&=&\sum_{l=1}^{\infty}G^{(l)}a^l=-\zeta_3a^2+\left(4\zeta_5+\frac{10}{3}\zeta_2\zeta_3\right)a^3+\ldots.
\end{eqnarray}
That is, one can expect that a similar structure of IR divergences in colour-ordered amplitudes 
holds for any planar gauge theory, not only $\mathcal{N}=4$ SYM. For a more detailed discussion and bibliography see, for example, \cite{Becher:2014oda} and references therein. I.e one can expect that the IR divergences also manifest themselves as $1/\epsilon^2$ and $1/\epsilon$ poles which will exponentiate \cite{Mueller:1979ih,Magnea:1990zb,Sterman:2002qn}. If the theory is also conformal the structure of IR divergences will be identical to the $\mathcal{N}=4$ SYM case:
\begin{equation}\label{LogMDimRegDiv}
\log M_n\Big{|}_{Div.} = -\sum_{l=1}^{\infty}\sum_{i=1}^n\frac{a^l}{4}\left(\frac{\Gamma^{(l)}_{cusp}}{(l\epsilon)^2}+\frac{G^{(l)}}{(l\epsilon)}\right)\left(\frac{\mu^2}{s_{ii+1}}\right)^{l\epsilon},
\end{equation}
but with (potentially) different $\Gamma_{cusp}(a)$ and $G(a)$.
It is usually assumed that $\Gamma_{cusp}(a)$ is the unique label of the theory.
Note that the kinematical dependence of the divergent part is consistent with the collinear factorisation
properties of the amplitudes, which may be considered as self-consistency conditions of the theory.

In $\mathcal{N}=4$ SYM, however, one can show \cite{BernBDS,Drummond:2007au} that not only the divergent parts of the amplitude are known to all loops, but also the finite parts $\mathcal{F}_n$ for $n=4,5$ are known up to a kinematically independent function of the coupling constant. They are given by ($s_{12}\equiv s$, $s_{23}\equiv t$ for $n=4$ case):
\begin{eqnarray}\label{FinPartDimReg}
\mathcal{F}_4&=&\frac{\Gamma_{cusp}(a)}{4}\log^2 \left(\frac{s}{t}\right)+c_4(a),\nonumber\\
\mathcal{F}_5&=&\frac{\Gamma_{cusp}(a)}{8}\sum_{i=1}^5\left(-\log\left(\frac{s_{ii+1}}{s_{i+1i+2}}\right) \log\left(\frac{s_{i-1i}}{s_{i+2i+3}}\right)\right)+c_5(a).
\end{eqnarray}
Here $c_{4,5}(a)$ are kinematically independent unknown functions of the coupling constant.
These relations are the consequences of the dual conformal symmetry (the anomalous Ward identities related to the dual conformal symmetry breaking via the presence of the IR regulator \cite{Drummond:2007au}). 

One can also rewrite relations (\ref{LogMDimReg}) and (\ref{FinPartDimReg}) in a slightly different way, which was first formulated in \cite{BernBDS} as the BDS ansatz:
\begin{eqnarray}\label{BDSExpDimReg}
M_n=\exp \left[\sum_{l=1}^{\infty}a^l\left(~f^{(l)}(\epsilon)~M_n^{(1)}(l\epsilon)+C^{(l)} \right) +E_n \right],
\end{eqnarray}
where: 
$f^{(l)}(\epsilon)=f^{(l)}_0+\epsilon f^{(l)}_1+\epsilon^2 f^{(l)}_2$ with
$f^{(l)}_0=\Gamma_{cusp}^{(l)}/2$, 
$f^{(l)}_1=lG^{(l)}/2$. 
The constants $f^{(l)}_2$, 
$C^{(l)}$ are unknown a priory but are related to the
kinematically independent contributions to the amplitude; $E_n$ is kinematically dependent but is  $O(\epsilon)$; $M_n(l\epsilon)$ is the one-loop normalised amplitude with a shifted parameter of the dimensional regularisation $\epsilon \mapsto
l\epsilon$. We want to stress that this relation is valid only for $n=4,5$. For $n>5$ the so called finite remainder correction appears \cite{Drummond:2007au,Drummond:2008aq}. This correction depends on the coupling constant and kinematical invariants only. Moreover, only such combinations of kinematical invariants, which are invariant under the so called dual conformal symmetry transformations (dual conformal cross-ratios) are allowed. This symmetry is the (super)conformal symmetry in the momentum space \cite{DrummondSuperconformalSymmetry} and its appearance is closely related to the integrability properties of the theory \cite{DrummondYangianSymmetry,Beisert:2010jq,Beisert:2010jr}.

We see that the all-loop answer for $n=4,5$ is, roughly speaking, given by the exponent of the one-loop amplitude times $\Gamma_{cusp}(a)$. The function $\Gamma_{cusp}(a)$ in turn satisfies the so called BES integral equation \cite{Beisert:2006ez,Eden:2006rx,Freyhult:2007pz}. As far as we know, there is no known solution to this equation in a closed form, but one can explicitly evaluate $\Gamma_{cusp}(a)$ up to any given order of perturbation in $a$ or $1/a$ using this equation \cite{Beisert:2006ez,Eden:2006rx,Basso:2009gh}. See Appendix \ref{a2} for details.

\subsection{Amplitudes on the Coulomb branch with massless external particles}\label{s22}
Amplitudes on the Coulomb branch of the theory are much less investigated in contrast to the previously discussed case. Probably the simplest way to consider the theory on the Coulomb branch \cite{HenrietteCoulomb} is to replace the gauge group $SU(N_c)$ with $U(N_c+M_c)$ and break it down to $U(N_c)\times U(M_c)$, where both $N_c$ and $M_c$ are considered large. The spectrum in such a theory is given by massive $W$ bosons and their superpartners which are $U(N_c)\times U(M_c)$ bi-fundamentals and a pair of massless $U(N_c)$ and $U(M_c)$ gluons $g$ with their corresponding superpartners \cite{HenrietteCoulomb}. More complicated gauge symmetry patterns are also possible \cite{HenrietteCoulomb}. Then there are two major possibilities: one can consider scattering amplitudes in a pure gluon sector where massive particles will appear only in loops and all on-shell particles are massless \cite{HennGiggs1}\footnote{More accurately, the authors of \cite{HennGiggs1} considered gauge symmetry broken down to $U(N_c) \times U(1)^{M_c}$, $N_c \gg M_c$. Then they considered the scattering of $U(1)^{M_c}$ states and the equal VEVs limit. This effectively corresponded to the scattering of massless particles, while massive particles propagated only in loops. The equal VEVs limit also restored the $U(N_c) \times U(M_c)$ gauge symmetry.} or one can consider scattering amplitudes in a $W$ boson sector where all on-shell particles are massive. 
\begin{figure}[t]
 \begin{center}
 %\leavevmode
  \epsfxsize=13cm
 \epsffile{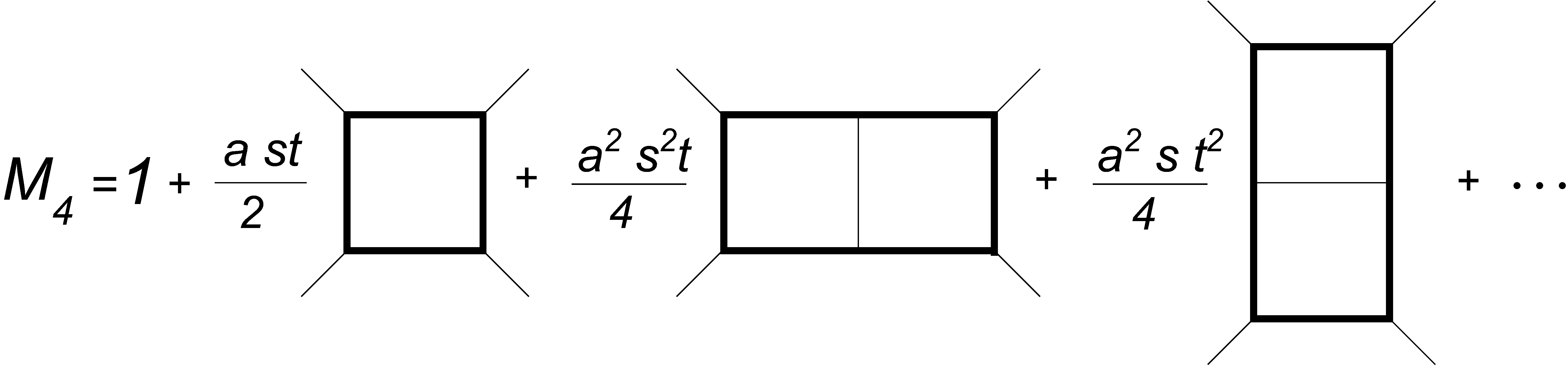}
 \end{center}\vspace{-0.2cm}
 \caption{Perturbative expansion of the four-point amplitude on the Coulomb branch with massless external particles \cite{HennGiggs1}. All masses are considered equal and small.  Thick black lines correspond to the massive propagators.}\label{fig1}
 \end{figure}

The first case, with external massless particles, was investigated in a series of papers \cite{HennGiggs1,HennGiggs2,HennGiggs3} for $n=4$ and $n=5$ colour-ordered amplitudes. See fig.\:\ref{fig1} for an example of the $n=4$ perturbative expansion. Direct computations suggest that the following pattern holds for the $M_n$ ratio\footnote{In the presence of massive particles the helicity structure of the amplitudes is more involved in comparison with the massless case \cite{HenrietteCoulomb}. Nevertheless, we assume that it is always possible to define an analogue of the tree-level MHV amplitude which will factorise.} if all masses of massive particles are equal:
\begin{equation}\label{LogMHigssg}
\log M_n = \sum_{i=1}^n\left(-\frac{\Gamma_{cusp}(a)}{8}\log^2\left( \frac{m^2}{s_{ii+1}} \right)
+\frac{\tilde{G}(a)}{2}\log\left( \frac{m^2}{s_{ii+1}} \right)\right)
+\mathcal{F}_n+O(m^2),
\end{equation}
where $m$ is the mass of $W$ bosons that runs in the loops and $\tilde{G}(a)$ is a (potentially) different collinear anomalous dimension compared to the massless case. The first term of the perturbative expansion is given by $\tilde{G}(a)=-\zeta_3a^2+\ldots$. We see that now IR divergences manifest themselves as powers of $\log(m^2)$, but their structure remains analogous to (\ref{LogMDimRegDiv}) from the dimensional regularisation case and is controlled by the same function $\Gamma_{cusp}(a)$:
\begin{equation}\label{LogMHigssgDiv}
\log M_n\Big{|}_{Div.} =\sum_{i=1}^n\left(-\frac{\Gamma_{cusp}(a)}{8}\log^2\left( \frac{m^2}{s_{ii+1}} \right)
+\frac{\tilde{G}(a)}{2}\log\left( \frac{m^2}{s_{ii+1}} \right)\right).
\end{equation}
The finite part of the amplitude $\mathcal{F}_n$ turns out to be very similar to the massless case for $n=4,5$ and is given by (\ref{FinPartDimReg}) with different kinematically independent contributions $c_{4,5}(a)$ \cite{HennGiggs1,HennGiggs3}. 

One can also write (\ref{LogMHigssg}) in a form similar to (\ref{BDSExpDimReg}) \cite{HennGiggs2}. For example, in the four-point case one can rewrite $\log M_4$ as:
\begin{eqnarray}\label{LogMHigssg1}
\log M_4 &=& -\frac{\Gamma_{cusp}(a)}{2} \log\left( \frac{m^2}{s} \right)\log\left( \frac{m^2}{t} \right)+\tilde{G}(a)\left( \log\left( \frac{m^2}{s} \right) + \log\left( \frac{m^2}{t} \right)\right)\nonumber\\
&+&\tilde{c}_4(a)+O(m^2),
\end{eqnarray}
where $\tilde{c}_4(a)$ is some kinematically independent  function of the coupling constant. Taking into account that the
one-loop correction is given by
\begin{equation}
M_4^{(1)}= \log\left( \frac{m^2}{s} \right)\log\left( \frac{m^2}{t} \right)-\frac{\pi^2}{2}+O(m^2),
\end{equation}
we see that, as in the massless case, the pattern that the all-loop result is given by the "exponent of the one-loop correction times Cusp anomalous dimension" holds:
\begin{eqnarray}\label{LogMHigssg3}
\log M_4 &=& -\frac{\Gamma_{cusp}(a)}{2}M^{(1)}_4+\tilde{G}(a)\left( \log\left( \frac{m^2}{s} \right) + \log\left( \frac{m^2}{t} \right)\right)\nonumber\\
&+&\tilde{c}_4(a)+O(m^2).
\end{eqnarray}
Note that in this case $\log M_4$ depends on two dimensionless parameters $m^2/s$ and $m^2/t$. This is not accidental and such a functional dependence is fixed by the dual conformal invariance which is not broken in this case \cite{HennGiggs1}. In this scenario, however, the dual conformal symmetry is interpreted as the symmetry of higher-dimensional space. Particle masses are interpreted as additional coordinates
in this space. So one can think of the scattering of massive particles in gauge theory with spontaneously broken
gauge symmetry as the scattering of massless states in higher-dimensional space \cite{HenrietteCoulomb}.

\subsection{Amplitudes on the Coulomb branch with massive external particles}
The scatting amplitudes in the $W$ boson sector have not been investigated at all until recently \cite{Caron-Huot:2021usw}. There, it was pointed out that the behaviour of the $W$ boson $n=4$ point amplitude $M_4$ can be linked to the behaviour of the correlation function $G_4$ of the 1/2-BPS operators with a large R-charge $K \gg 1$ \cite{Caron-Huot:2021usw}:
\begin{equation}\label{Korr}
G_4=\langle \mathcal{O}_1(x_1)\ldots \mathcal{O}_4(x_4) \rangle,
\end{equation}
where $\mathcal{O}_1=\mbox{Tr}(\bar{X}^{2K})$, $\mathcal{O}_2=\mbox{Tr}(X^K\bar{Z}^K)$, $\mathcal{O}_3=\mbox{Tr}(Z^{2K})$ and $\mathcal{O}_4=\mbox{Tr}(Z^K\bar{X}^K)$; $X$ and $Z$ are two of three $\mathcal{N}=4$ SYM complex scalar fields labelled as $X,Y,Z$. It was shown in \cite{Hex1,Basso:2017khq,Coronado:2018cxj,Caron-Huot:2021usw} that $G_4$ can be represented as a square of the so called octagon function $\mathbb{O}_0$:
\begin{equation}\label{Korr4}
G_4=\frac{\mathbb{O}_0^2}{(x_{12}^2x_{23}^2x_{34}^2x_{41}^2)^K} ,
\end{equation}
In the small equal mass limit (which translates into the light cone limit for the correlator $G_4$ as $x_{ii+1}^2\equiv m^2 \ll 1$ ) the following relation is expected to hold \cite{Caron-Huot:2021usw}:
\begin{equation}\label{LogM4W}
\log M_4 = \log \mathbb{O}_0+O(m^2).
\end{equation}
The octagon function $\mathbb{O}_0$ in its turn can be represented as a determinant of the infinite-dimensional matrix:
\begin{eqnarray}\label{OctDef1}
\mathbb{O}_0&=&\det (1-\mathbb{K}_0),\nonumber\\
(\mathbb{K}_0)_{nm}&=&\sum_{l=n+m-1}^{\infty}C_{nm}^{(l)}~(-a/2)^l~F_{l}(z,\bar z),
\end{eqnarray}
where the coefficients $C_{nm}^{(l)}$ are given by:
\begin{equation}
C_{nm}^{(l)}=\frac{-(2m-1)(2l)!(l-1)!l!}{(l-(n+m-1))!(l+(n+m-1))!(l-|n+m|)!(l+|n+m|)!},
\end{equation}
and $F_{l}(z,\bar z)$ are the $l$-loop Ussyukina-Davydychev box functions \cite{Usyukina:1992jd,Usyukina:1993ch} written in terms of the conformal cross-ratios $z$ and $\bar z$:
\begin{equation}
F_{l}(z,\bar z)=(-1)\sum_{j=l}^{2l}\frac{j!(-1)^{j}\log^{2l-j}\left(\frac{ z \bar{z} }{(z-1)(\bar z -1)} \right)}{l!(j-l)!(2l-j)!}
\left( 
\frac{\mbox{Li}_{l}\left(\frac{z}{z-1}\right) - \mbox{Li}_{l}\left(\frac{\bar z}{\bar z-1}\right)}
{z-\bar z} 
\right).
\end{equation}
In terms of the Mandelstam invariants $s$, $t$ and the masses of the external particles ($m_i^2=p_i^2$) variables $z$, $\bar z$ are given by:
\begin{equation}
\frac{m_1^2m_3^2}{st}=\frac{z\bar z}{(1-z)(1- \bar z)},~\frac{m_2^2m_4^2}{st}=\frac{1}{(1-z)(1- \bar z)}.
\end{equation}
In the case under consideration, all masses are equal $m_1^2=\ldots=m_4^2\equiv m^2$. Let us stress that here we are considering the connection (duality) between the correlation functions in the massless $\mathcal{N}=4$ SYM (at the origin of the moduli space) and the amplitudes with massive particles in the $\mathcal{N}=4$ SYM with spontaneously broken gauge symmetry in the small mass limit. Relation (\ref{LogM4W}) is also similar to the well-known duality between the $\mathcal{N}=4$ SYM planar amplitudes \cite{Eden:2010zz} in the pure massless case and the correlation functions of the operators from the stress-tensor supermultiplet.

The determinant (\ref{OctDef1}) can be bootstraped \cite{Coronado:2018cxj}, which allows one to obtain an explicit expression for $\log M_4$ in the small mass limit (there is also an alternative approach based on integrable differential equations \cite{Belitsky:2019fan,Belitsky:2020qrm}\footnote{The bootstrap approach \cite{Coronado:2018cxj} allows one to fix the kinematical dependence of $\log M_4$ while the explicit expressions for the kinematically independent parts of $\log M_4$ to our knowledge were first obtained in \cite{Belitsky:2019fan,Belitsky:2020qrm}.}):
\begin{equation}\label{LogM4W1}
\log M_4 = -\frac{\Gamma_{oct}(a)}{4}\log^2 \left( \frac{m^4}{st} \right)-\frac{D(a)}{2}+O(m^2),
\end{equation}
where the functions $\Gamma_{oct}(a)$ and $D(a)$ can be written explicitly in terms of elementary functions \cite{Belitsky:2019fan,Belitsky:2020qrm}:
\begin{eqnarray}\label{GammaOctSeries}
\Gamma_{oct}(a)&=&\frac{2}{\pi^2}\log \left( \cosh \left(\pi \sqrt{2a}\right)\right)=2a-4\zeta_2a^2+32\zeta_4a^3+\ldots,\nonumber\\
D(a)&=&\frac{1}{4}\log \left(\frac{\sinh( 2\pi \sqrt{2a} )}{2\pi \sqrt{2a}} \right)=2\zeta_2 a-8\zeta_4 a^2-\frac{128\zeta_6}{3}a^3+\ldots.
\end{eqnarray}
We see that in this case the IR behaviour is \emph{not} controlled by $\Gamma_{cusp}(a)$ \cite{Caron-Huot:2021usw}, but in terms of dependence on kinematical invariants this is still "the exponent of the one-loop correction", because
$M^{(1)}_4$ is proportional to the one-loop Ussyukina-Davydychev box function which in the small mass limit is given by:
\begin{eqnarray}
M^{(1)}_4=\frac{1}{2}\log^2\left( \frac{m^4}{st} \right)+\frac{\pi^2}{6}+O(m^2).
\end{eqnarray}
Using this relation, we can write (\ref{LogM4W1}) in a form identical to (\ref{LogMHigssg3}) but with $\tilde{G}(a)=0$ and a different value of the kinematically independent part:
\begin{equation}\label{LogM4W2}
\log M_4 = -\frac{\Gamma_{oct}(a)}{2}M^{(1)}_4+\tilde{c}_4(a)+O(m^2).
\end{equation}
Here $\tilde{c}_4(a)$ is a different function compared to (\ref{LogMHigssg3}).
The dependence of $\log M_4$ on a single argument $m^4/st$ is fixed by the conformal symmetry
of the correlation function which can be interpreted as dual conformal symmetry of the massive amplitude.

Using a simple identity
\begin{equation}
\log^2 \left( \frac{m^4}{st} \right)=2\log^2 \left( \frac{m^2}{s} \right)+2\log^2 \left( \frac{m^2}{t} \right)-\log^2 \left( \frac{t}{s} \right),
\end{equation}
we can rewrite (\ref{LogM4W1}) in a form similar to (\ref{LogMHigssg}):
\begin{equation}\label{LogM4W3}
\log M_4 = -\frac{\Gamma_{oct}(a)}{2}\left( \log^2 \left( \frac{m^2}{s} \right)+\log^2 \left( \frac{m^2}{t} \right) \right)+\tilde{\mathcal{F}}_4+O(m^2).
\end{equation}
We see that the divergent part of the amplitude is given by (\ref{LogMHigssgDiv}) with $\Gamma_{cusp}(a)\mapsto 2\Gamma_{oct.}(a)$ and $\tilde{G}(a)\mapsto 0$ replacements,
with the finite part $\tilde{\mathcal{F}}_4$ given by:
\begin{equation}
\tilde{\mathcal{F}}_4=\frac{\Gamma_{oct}(a)}{4}\log^2\left( \frac{s}{t} \right)-\frac{D(a)}{2}.
\end{equation}

It is also important to note that $\Gamma_{oct}(a)$ and $\Gamma_{cusp}(a)$ are in fact related functions which satisfy the same generalized BES equation. As the solutions of the same equation they are related by different values of an interpolation parameter \cite{Basso:2020xts}. Numerically, they are still different, which is very interesting in the light of the common belief \cite{ConformalProperties4point} that all IR properties of the amplitudes are controlled by
the same function $\Gamma_{cusp}$.  

It is also interesting to consider higher-point amplitudes of $W$ bosons in the small mass limit and investigate their IR properties. The $n=5$ point amplitude is the first natural candidate.

\section{IR properties of four- and five-point $W$ boson amplitudes in $\mathcal{N}=4$ SYM}\label{s3}
\subsection{$n=4$ point amplitude}
\begin{figure}[t]
 \begin{center}
 %\leavevmode
  \epsfxsize=13cm
 \epsffile{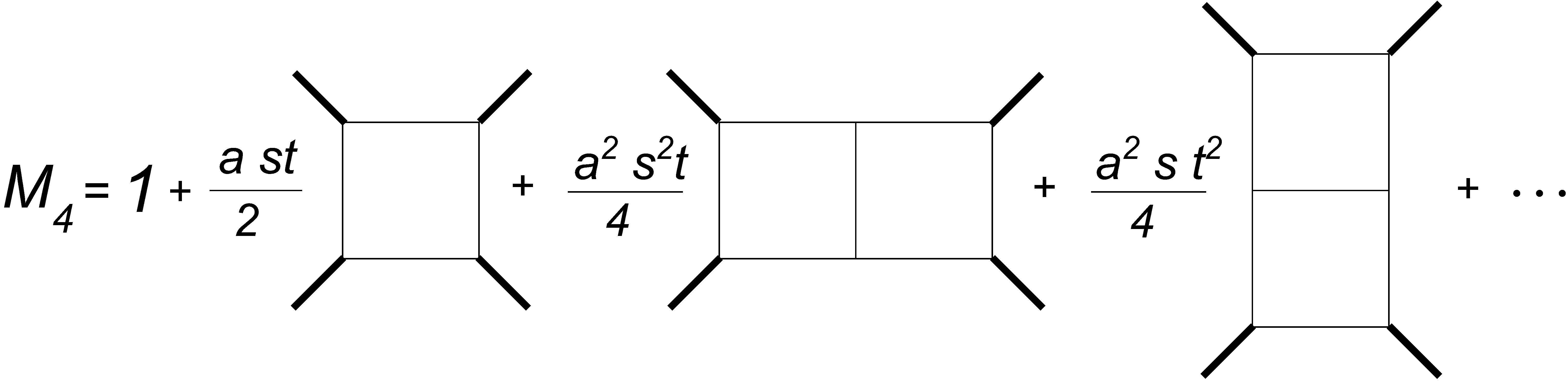}
 \end{center}\vspace{-0.2cm}
 \caption{Perturbative expansion of the four-point $W$ boson amplitude on the Coulomb branch with massive external particles. All masses are considered equal and small. Thick black lines correspond to the massive external states.}\label{fig2}
 \end{figure}
Before considering the $n=5$ amplitude it is instructive to see how (\ref{LogM4W1}) arises from perturbative computations from the "amplitude point of view". Let us remind that we are considering the $n=4$ point amplitude of $W$ bosons on the Coulomb branch. All masses of $W$'s are taken small and equal to each other.
The dual description of the amplitude in terms of the octagon (i.e. in terms of the four point correlation function) suggests that the perturbative expansion for the amplitude is given by the same master integrals as in the massless case but with external momenta squared equal to the square of ($p_i^2=m^2$). Hereafter we will assume that this is a correct representation of the amplitude in the small mass limit.
This is in fact identical to the "off-shell" regularisation \cite{ConformalProperties4point} considered previously in the literature (see fig.\:\ref{fig2}). This approach was obscured at that time since it was not obvious that such an ad hoc extension of the massless results will be consistent with gauge-invariance. It would be interesting to faithfully reproduce this expansion from unitarity cuts using the tree-level amplitudes on the Coulomb branch from \cite{HenrietteCoulomb} or similar results. Note that to faithfully reproduce an expansion with purely massless internal lines, one will likely require a more involved gauge symmetry breaking pattern than the one described at the beginning of section \ref{s21}.

An additional justification for the aforementioned form of the loop expansion of the four-point amplitude can be given by the fact that at the level of integrands\footnote{To reproduce the expansion from fig.\:\ref{fig2}, one has to use dual variables and rewrite each propagator $x_{li}^2$ in the $D$-dimensional integrand as a sum of $D=4$ and extra dimensional parts $x_{li}^2=x_{li}^{2,D=4}+y_{il}^2$. Then one can impose a $\delta^{D-4}(y_l)$ constraint on the coordinates of integration vertices together with a light-cone constraint on dual coordinates associated with external momenta $y_i^2=0$ \cite{Caron-Huot:2021usw}. These constraints from the $D=4$ point of view will give us massless propagators and massive external lines with $m^2=-y_{ii+1}^2$. $D$ dimensional massless momenta with dual coordinates $x_{ii+1}^2=0$ are interpreted as massive four dimensional ones.} the four-point amplitude divided by the tree-level one is identical among $D=4$ $\mathcal{N}=4$ SYM, $D=6$ $\mathcal{N} = (1,1)$, $D=8$ $\mathcal{N}=1$ SYM and $D=10$ $\mathcal{N} = 1$ SYM theories \cite{Boels:2012ie,Mafra:2015mja,Mafra:2008ar}. Integrands of the amplitudes in massless theories in  higher dimension can be interpreted as integrands of the massive amplitudes in lower dimension with spontaneously broken gauge symmetry (for example, see discussion in \cite{HenrietteCoulomb}).

\begin{figure}[t]
 \begin{center}
 %\leavevmode
  \epsfxsize=13cm
 \epsffile{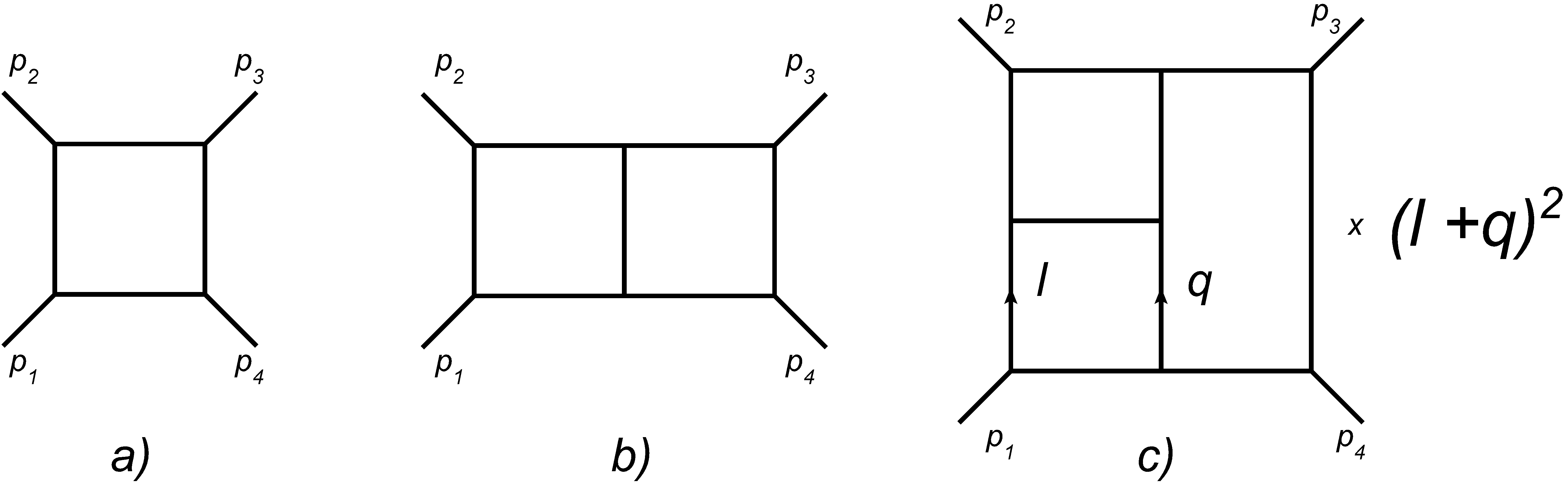}
 \end{center}\vspace{-0.2cm}
 \caption{Integrals contributing to $M_4^{(l)}$ up to three loops. a) is the $B$ Box integral, b) is the $DB$ Double Box integral and c) is $TC$ the "Tennis Court" one. Here all external momenta $p_i$ are considered massive $p_i^2=m^2$.}\label{fig3}
 \end{figure}

So we conclude that up to three loops in the small mass limit $M_4^{(l)}$ are given by the following well-known combinations of scalar integrals\footnote{Hereafter we hide $(-1)$ factors from $M_n^{(l)}$ definitions into $a\mapsto -a$ replacement in the perturbative expansion. This is consistent at the orders of perturbation theory we are working with.}:
\begin{equation}\label{M4exp1}
M^{(1)}_4=\frac{1}{2}~st~B(s,t)+O(m^2),~
M^{(2)}_4=\frac{1}{4}\left( s^2t~DB(s,t)+st^2~DB(t,s) \right)+O(m^2),\\
\end{equation}
and
\begin{equation}\label{M4exp2}
M^{(3)}_4=\frac{1}{8}\left( s^3t~TB(s,t)+st^3~TB(t,s)+2s^2t~TC(s,t)+2ts^2~TC(t,s) \right)+O(m^2)
\end{equation}  
where $B$, $DB$ and $TB$ are respectively, the Box integral, the Double Box integral and the Triple Box integrals; $TC$ is the so called "Tennis Court" integral. All integrals are represented in fig.\:\ref{fig3}. We explicitly show the dependence on the Mandelstam invariants $s$ and $t$ and suppress the dependence on $m^2$. 
We also neglect possible $m^2$ dependences in the numerators since these are $O(m^2)$ corrections. All these integrals are dual conformally invariant. 
%which from "the correlation function point of view" is the reflection of ordinary conformal symmetry 

All the $B$, $DB$, $TB$ and $TC$ integrals with massless internal propagators and massive external legs, both in the $s$ and $t$ channels when multiplied by the corresponding powers of $s$ and $t$ from (\ref{M4exp1}) and (\ref{M4exp2}) are equal to the Ussyukina-Davydychev box functions $F_l(z,\bar z)$ due to the so called "magical identities" \cite{Drummond:2006rz} between the four-point dual conformally invariant integrals. 
%More explicitly this gives us:
%\begin{eqnarray}
%&&st~B(s,t)=F_1(x,x),\nonumber\\
%&&s^2t~DB(s,t)=st^2~DB(t,s)=F_2(x,x),\nonumber\\
%s^3t~TB(s,t)+st^3~TB(t,s)
%\begin{eqnarray}
%where 
For our purpose we find it more convenient to use the $x=p_1^2p_3^2/st$, $y=p_2^2p_4^2/st$ variables instead $z$, $\bar z$. We give an explicit expression for $F_l(x,y)$ in Appendix \ref{a1}. This allows us to rewrite $M^{(l)}_4$ up to three loops simply as:
\begin{equation}
M^{(1)}_4=\frac{1}{2}F_1(x,x)+O(m^2),
~ M^{(2)}_4=\frac{1}{2}F_2(x,x)+O(m^2),
~M^{(3)}_4=\frac{3}{4}F_3(x,x)+O(m^2),
\end{equation}
where $x=m^4/st$. These are exactly the first three terms of the perturbative expansion of the octagon $\mathbb{O}_0$ (\ref{OctDef1}). Small 
$m^2$ expansions of $F_l(x,x)$ up to $l=3$ are given by:
\begin{eqnarray}\label{BoxSmallMassExp}
F_1(x,x)&=&\log^2(x)+2\zeta_2+O(m^2),\nonumber\\
F_2(x,x)&=&\frac{\log^4(x)}{4}+\frac{3\zeta_2}{2}\log^2(x)+\frac{21\zeta_4}{2}+O(m^2),\nonumber\\
F_3(x,x)&=&\frac{\log^6(x)}{36}+\frac{5\zeta_2}{6}\log^4(x)
+\frac{35\zeta_4}{2}\log^2(x)+\frac{4\zeta_6}{155}+O(m^2).
\end{eqnarray}
Note the absence of $\log^{2n+1}$ terms, which implies the absence of $\tilde{G}(a)$ in the amplitude. Substituting this expressions in $\log M_4$:
\begin{eqnarray}
\log M_4&=&-a~M_4^{(1)}+
a^2\left( M_4^{(2)}-\frac{1}{2}\left(M_4^{(1)}\right)^2 \right)\nonumber\\
&-&a^3\left(\frac{1}{3}\left(M_4^{(1)}\right)^3 -M_4^{(1)}M_4^{(2)}+M_4^{(3)} \right)+\ldots
\end{eqnarray}
%\begin{equation}
%\log M_4=a~M_4^{(1)}+
%a^2\left( M_4^{(2)}-\frac{1}{2}\left(M_4^{(1)}\right)^2 \right)+
%a^3\left(\frac{1}{3}\left(M_4^{(1)}\right)^3 -M_4^{(1)}M_4^{(2)}+M_4^{(3)} \right)+\ldots
%\end{equation}
we obtain:
\begin{equation}
\log M_4=\left(-\frac{a}{2}+\zeta_2a^2-8\zeta_4a^3+\ldots \right)\log^2(x)+\left(-\zeta_2a+4\zeta_4a^2-\frac{64}{3}\zeta_6a^3+\ldots \right)+O(m^2),
\end{equation}
which is exactly (\ref{LogM4W1}) expanded up to $a^3$. 

Note that one can use the following strategy of computations \cite{ConformalProperties4point}. One can only compute the divergent part of the amplitude (see $\log^2(m^2)$ terms in (\ref{LogM4W3})) and then take into account the dual conformal symmetry (the fact that the whole $\log M_4$ must depend only on $m^4/st$). This will uniquely fix the finite part $\tilde{\mathcal{F}}_4$ up to a kinematically independent additive constant. In the $n=4$ point case, this observation is trivial due to a simple structure of loop integrals at least in the first several loop orders but will be useful in the $n=5$ case. 

In conclusion, let us also compare the Regge limits of the four-point $W$ boson amplitude considered here
and the four-point amplitude (\ref{LogMHigssg1}) that is Regge exact and can be rewritten in the Regge form without additional assumptions \cite{HennGiggs2}: 
\begin{equation}\label{ReggeForm}
M_4=\beta(t)\left(\frac{s}{m^2}\right)^{\alpha(t)-1}+O(m^2),
\end{equation}
where:
\begin{equation}
\beta(t)=\exp\left[ -\tilde{G}(a)\log\left( \frac{t}{m^2} \right)+c_4(a) \right],
\end{equation}
and the Regge trajectory $\alpha(t)$ is given by:
\begin{equation}
\alpha(t)=1-\frac{\Gamma_{cusp}(a)}{2}\log\left( \frac{t}{m^2} \right)-\tilde{G}(a).
\end{equation}
In our case of $W$ boson scattering the amplitude 
can also be written in a similar form to some extent. Considering $s \gg t$, which in dimensionless variables translates
into $\log(m^2/t) \gg \log(m^2/s)$ we can neglect $\log^2(m^2/s)$ in the log of the amplitude (\ref{LogM4W1}).  In the exponent, strictly speaking, we can do so only if $\log(m^2/s) \ll 1$. This formally gives us the amplitude written in the form of (\ref{ReggeForm}) but with: 
\begin{equation}
\beta(t)=\exp\left[ -\frac{\Gamma_{oct}}{4}\log^2\left( \frac{m^2}{t} \right)-\frac{D(a)}{2} \right],
\end{equation}
and the Regge trajectory $\alpha(t)$ is given by:
\begin{equation}
\alpha(t)=1-\frac{\Gamma_{oct}(a)}{2}\log\left( \frac{t}{m^2} \right).
\end{equation}
Note that we have $\alpha(t=m^2)=1$ as expected. However, the $\log(m^2/s) \ll 1$ condition is likely inconsistent with approximations made in derivation of (\ref{LogM4W1}).

\subsection{$n=5$ point amplitude} 
\begin{figure}[t]
 \begin{center}
 %\leavevmode
  \epsfxsize=14cm
 \epsffile{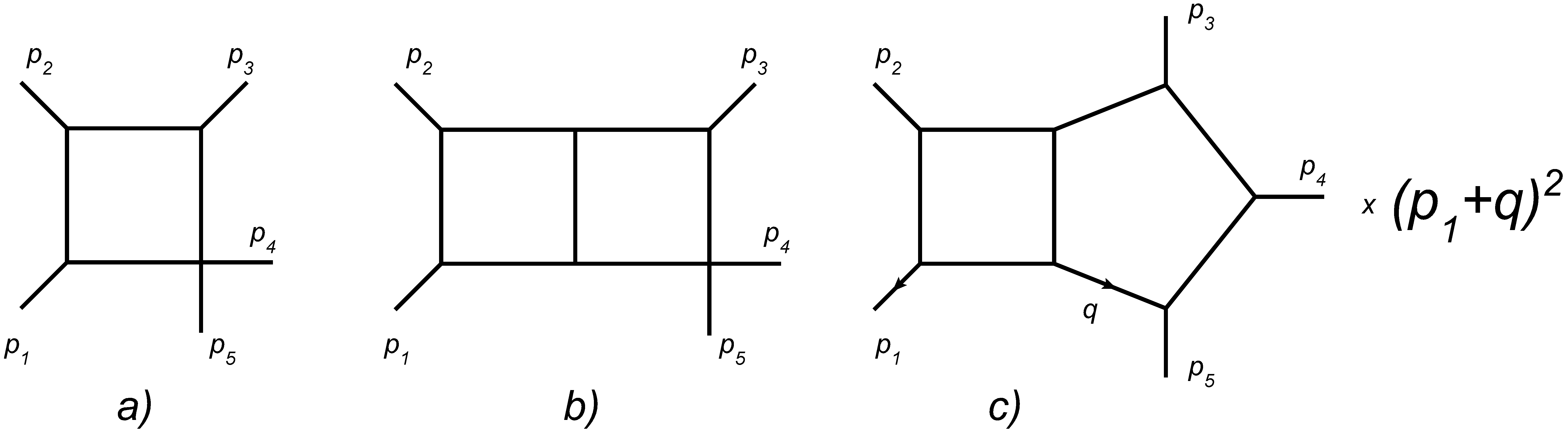}
 \end{center}\vspace{-0.2cm}
 \caption{Integrals contributing to $M_5^{(l)}$ at one and two loops. a), b) are the $B$ and $DB^{(h)}$ integrals for five-point kinematics. c) is the $PB$ PentaBox integral. Here, as in the previous case, all external momenta $p_i$ are considered massive $p_i^2=m^2$.}\label{fig4}
 \end{figure}
In the massless case the structure of the expansion of the five-point amplitude is rather
involved due to the presence of the integrals proportional to $\epsilon$ \cite{Bern:2006vw}. Such integrals are important for the exponentiation structure (\ref{BDSExpDimReg}) due to the $\epsilon \times 1/\epsilon$ interference. The five-point amplitude can be split into odd and even parts, and all proportional to $\epsilon$ integrals are condensed in the odd part\footnote{It is also worth mentioning that up to $O(\epsilon)$ the contribution from the odd part drops out from $\log M_5$ up to two loops \cite{Bern:2006vw,Eden:2010zz}.}.
We assume that the analogues of the integrals proportional to $\epsilon$ can be neglected in our case, which is identical to the conjecture made in \cite{HennGiggs3}. Thus, we assume that for $l=1,2$ the five-point amplitude of $W$ bosons divided by the tree-level one in the small mass limit can be written as:
\begin{eqnarray}\label{M5int}
M^{(1)}_5&=&\frac{\mathbb{P}_5}{4}\left( s_1s_2~B \right)+O(m^2),\nonumber\\
M^{(2)}_5&=&\frac{\mathbb{P}_5}{8}\left( s_1^2s_2~DB^{(h)}+s_1 s_2^2~DB^{(v)}+s_2 s_3 s_5~PB \right)+O(m^2),
\end{eqnarray} 
where $B$ is the one-loop Box integral with five consecutively ordered external momenta represented in fig.\:\ref{fig4}, $DB^{(h),(v)}$ are the horizontal and vertical Double Box integrals with five consecutively ordered external momenta represented in fig.\:\ref{fig4}, 
 $PB$ is a PentaBox integral with the numerator shown in fig.\:\ref{fig4}, $s_i\equiv (p_i+p_{i+1})^2$ and 
\begin{equation}
\mathbb{P}_5=\sum_{n=0}^4\mathbb{P}^n,
\end{equation}
where $\mathbb{P}$ is the operator that shifts all labels of external momenta by $+1$, $\mbox{mod}(5)$. For example, $\mathbb{P}^2\log (s_1/s_5)=\log (s_3/s_2)$. All integrals depicted in fig.\:\ref{fig4} are also dual conformally invariant.

As in the previous case of the $n=4$ point amplitude an additional justification that (\ref{M5int}) correctly reproduces the five-point amplitude in the small mass limit can be found from computations of the integrands in $D=10$ $\mathcal{N}=1$ SYM theory \cite{Mafra:2008gkx,Mafra:2015mja}. One can see \cite{Mafra:2015mja} that the two-loop integrand is given by the same integrals as in (\ref{M5int}) but with different numerators.
These numerators in the four-dimensional kinematical limit reduce to those in (\ref{M5int}). The one-loop five-point amplitude in \cite{Mafra:2015mja} is given by the Box and Pentagon integrals with the numerator \cite{Mafra:2015mja}. In the four-dimensional kinematical limit these integrals once again reduce to the sum of Box integrals from (\ref{M5int}) plus $O(m^2)$ corrections. Nevertheless it would be interesting to try to directly derive (\ref{M5int}) from $D=4$ $\mathcal{N}=4$ SYM on the Coulomb branch by means of unitarity cuts using the results of \cite{HenrietteCoulomb} or similar ones, or even an explicit Feynman diagram computation.

Returning to our perturbative expansion (\ref{M5int}), we see that the Box integrals are also proportional to the Ussyukina-Davydychev box functions but now with different values of $x$ and $y$. This allows us to write $M_5^{(1)}$ as:
\begin{eqnarray}\label{M51loop}
M^{(1)}_5&=&\frac{\mathbb{P}_5}{4}F_1\left( \frac{m^2s_4}{s_1s_2},\frac{m^4}{s_1s_2} \right)=
\frac{\mathbb{P}_5}{4}\left(\log\left( \frac{m^2s_4}{s_1s_2} \right)\log\left( \frac{m^4}{s_1s_2} \right)+2\zeta_2\right)+O(m^2)\nonumber\\
&=&\sum_{i=1}^5\frac{1}{2}\log^2\left( \frac{m^2}{s_i} \right)+\sum_{i=1}^5\frac{1}{4}\log\left(\frac{s_{i}}{s_{i+1}}\right) \log\left(\frac{s_{i-1}}{s_{i+2}}\right)+\frac{5\zeta_2}{2}+O(m^2),
\end{eqnarray} 
which is consistent with general considerations regarding the structure of IR divergences and suggests an exponentiation pattern:
\begin{equation}\label{logM5conject}
\log M_5=-\frac{\Gamma_{oct}(a)}{2}M_5^{(1)}+\tilde{c}_5(a)
\end{equation}
similar to the $n=4$ case. To evaluate the two-loop correction and verify the conjecture (\ref{logM5conject}), we restrict ourselves to the single scale $s_i=Q^2$, $i=1,\ldots,5$ kinematical limit. This allows us to avoid working with the full $PB$ integral in general kinematics. We are going to address the evaluation of the
PentaBox integral in general kinematics in a separate upcoming publication \cite{futur}.

In the single scale $s_i=Q^2$ kinematical point all integrals can be more or less straightforwardly evaluated by the
Mellin-Barnes representation technique \cite{Smirnov:2004ym,Smirnov:2006ry}. This gives us ($t=m^2/Q^2$, we hope that will be no confusion with the Mandelstam variable):
\begin{equation}\label{Bint}
s_1s_2~B\Big{|}_{s_i=Q^2}=F_1(t,t^2)=2\log^2(t)+2\zeta_2+O(m^2),
\end{equation}
\begin{eqnarray}\label{DBint}
s_1s_2^2~DB^{(h)}\Big{|}_{s_i=Q^2}&=&
F_2(t,t^2)=\log^4(t)+\frac{13\zeta_2}{2}\log^2(t)+\frac{21\zeta_4}{2}+O(m^2),\nonumber\\
s_3^2s_4~DB^{(v)}\Big{|}_{s_i=Q^2}&=&
F_2(t,t^2)=\log^4(t)+\frac{13\zeta_2}{2}\log^2(t)+\frac{21\zeta_4}{2}+O(m^2),
\end{eqnarray}
as well as:
\begin{equation}\label{PBint}
s_2 s_3 s_5~PB\Big{|}_{s_i=Q^2}=3\log^4(t)+5\zeta_2\log^2(t)+5\zeta_4+O(m^2).
\end{equation}
The results for the $DB$ integrals agree with the Ussyukina-Davydychev box functions expansion. We have also performed numerical checks of all our results using the FIESTA mathematica package \cite{SmirnovFIESTA,Smirnov:2021rhf}. See Appendix \ref{a1} for details.

Substituting (\ref{Bint}), (\ref{DBint}) and (\ref{PBint}) in the $\log M_5$ perturbative expansion 
\begin{equation}
\log M_5=-a~M_5^{(1)}+
a^2\left( M_5^{(2)}-\frac{1}{2}\left(M_5^{(1)}\right)^2 \right)+\ldots,
\end{equation}
we have
\begin{equation}
\log M_5\Big{|}_{s_i=Q^2}=5\left( -\frac{a}{2}+\zeta_2a^2+\ldots \right)\log^2(t)+\tilde{C}(a)+O(m^2),
\end{equation}
where $\tilde{C}(a)$ is $t$-independent.
The coefficient before $\log^2(t)$ agrees with the $\Gamma_{oct}(a)$ two-loop expansion:
\begin{equation}
-\frac{5}{4}\Gamma_{oct}(a)=5\left(-\frac{a}{2}+\zeta_2a^2+\ldots\right).
\end{equation}
This, together with with the one-loop result (\ref{M51loop}), suggests that the divergent part of the amplitude is given by
\begin{equation}\label{M5div}
\log M_5\Big{|}_{Div}=-\frac{\Gamma_{oct}(a)}{4}\sum_{i=1}^5\log^2\left( \frac{m^2}{s_i} \right),
\end{equation}
which can be considered as the main result of this article. We see that, as in the four-point case, the IR divergent part of the amplitude is likely controlled by $\Gamma_{oct}(a)$.
This, in turn, with the requirement that the functional dependence of $\log M_4$ is given by
\begin{equation}\label{CR5p}
\frac{m^2s_4}{s_1s_2},~\frac{m^4}{s_1s_2},\ldots,\mathbb{P}^4 \left( \frac{m^2s_4}{s_1s_2} \right),~\mathbb{P}^4\left( \frac{m^4}{s_1s_2} \right),
\end{equation}
implies that the finite part $\tilde{\mathcal{F}}_5$ (up to two loops) must be equal to:
\begin{equation}\label{FinPart5}
\tilde{\mathcal{F}}_5=-\frac{\Gamma_{oct}(a)}{8}\sum_{i=1}^5\log\left(\frac{s_{i}}{s_{i+1}}\right) \log\left(\frac{s_{i-1}}{s_{i+2}}\right)+\tilde{c}_5(a).
\end{equation}
Here $\tilde{c}_5(a)$ is an independent constant. List (\ref{CR5p}) can be considered as a dual conformal invariance constraint and can be deduced from the structure of conformal cross-ratios for $n=5$ points in the small mass limit.

The form of the finite part (\ref{FinPart5}) is in agreement with the (\ref{logM5conject}) conjecture. In other words, roughly speaking, if "the $\log(m^2)$ terms exponentiate, all $M_5$ will also exponentiate since all arguments on which $M_5$ depends are proportional to $m^2$".

\begin{figure}[t]
 \begin{center}
 %\leavevmode
 \epsfxsize=6cm
\epsffile{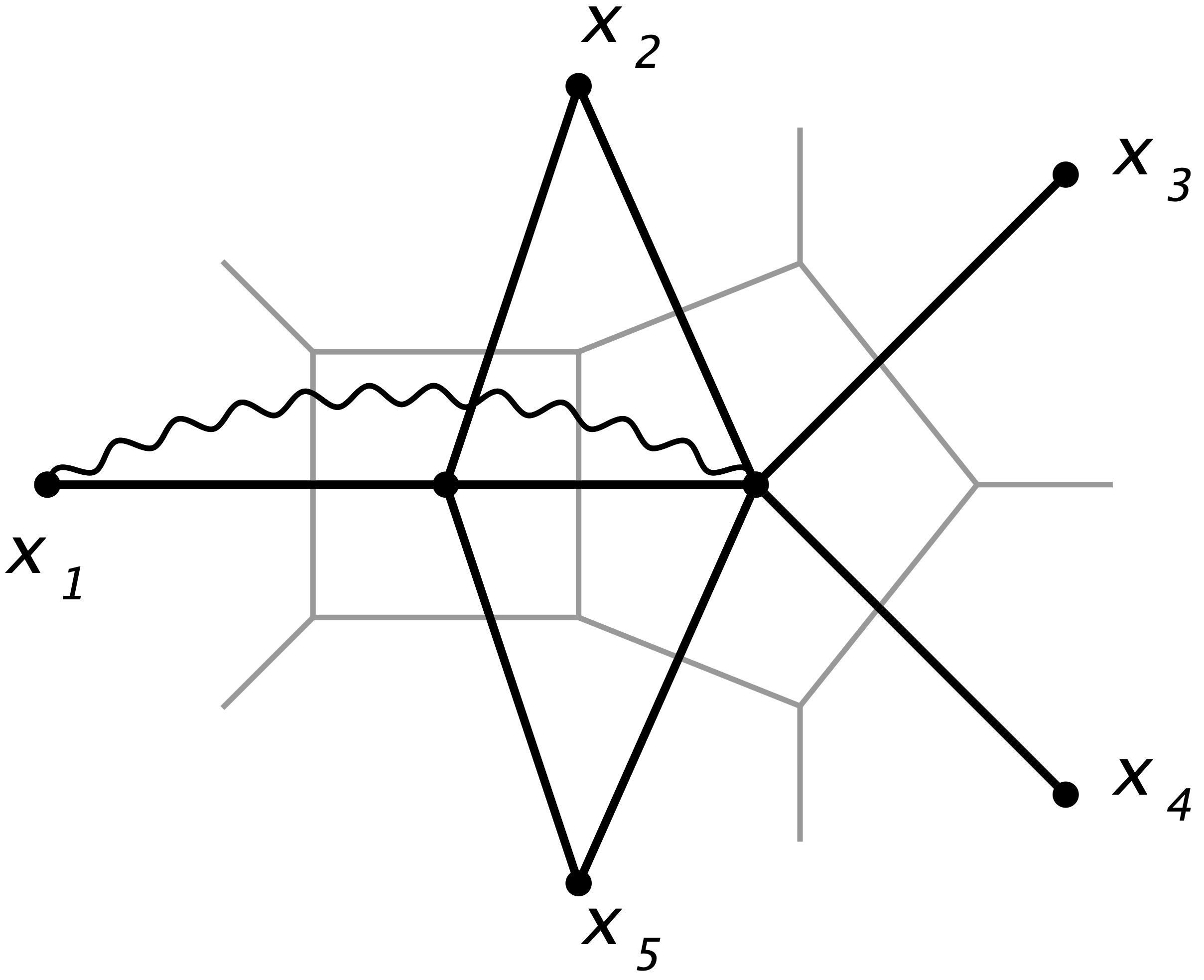}
\end{center}\vspace{-0.2cm}
\caption{PentaBox contributing to the five-point correlation function of $1/2$-BPS operators with a large $R$-charge. The integral is written in the dual coordinates. A wavy line corresponds to the numerator. 
%The numeration of $x_i$ coordinates points is taken from.
}\label{fig5}
\end{figure}

In conclusion, let us discuss the situation with $n \geq 5$ amplitudes. From the (dual) conformal symmetry one can see that if the five-point amplitude can be represented by dual conformal master integrals with massless propagators and massive external lines, it must depend on five cross-ratios 
\begin{equation}\label{CR}
\frac{x_{12}^2x_{43}^2}{x_{14}^2x_{23}^2},~\frac{x_{14}^2x_{23}^2}{x_{13}^2x_{42}^2},
~~\frac{x_{51}^2x_{34}^2}{x_{13}^2x_{45}^2},~\frac{x_{51}^2x_{43}^2}{x_{14}^2x_{53}^2},
~~\frac{x_{12}^2x_{45}^2}{x_{14}^2x_{25}^2}
\end{equation}
plus cross ratios generated from this set by cyclic shifts of $x_{ij}^2$ labels (action of the $\mathbb{P}^n$ operators with $n=1,\ldots,4$) \cite{Dectagon}. For example, the PentaBox integral from fig.\:\ref{fig5} depends on these five cross-ratios (\ref{CR}). One can see that all these cross-ratios in the small mass limit $x_{ii+1}^2=p_i^2$, $p_i^2=m^2\ll 1$ are proportional to $m^2$ or $m^4$. We also see that in the small mass limit one can indeed choose (\ref{CR5p}) as a set of independent variables for the amplitude. That is our small mass limit is equivalent to the small cross-ratio limit for $n=4,5$.

In the $n=6$ and higher-point cases, conformal cross-ratios, which are $m^2$ independent, appear. For general kinematics they are not small in the $p_i^2=m^2\ll 1$ limit. In the simplest $n=6$ case, there will be three such cross-ratios $u_{1,2,3}$. Parts of the amplitude which will depend on $u_{1,2,3}$ will not be constrained by possible $\log (m^2)$ terms exponentiation. This is equivalent to the appearance of the reminder 
function in the pure massless case of the BDS ansatz \cite{Drummond:2007au}. However, as was recently shown  \cite{Basso:2020xts} for the $n=6$ case, the reminder function will be
greatly simplified and can be described by a closed expression if one considers the small cross-ratio $u_{1,2,3} \ll 1$ limit.  So one can hope that there will be significant simplifications in the "all cross-ratios are small" limit for correlation functions and massive amplitudes, and closed all-loop answers can be obtained for $n \geq 6$. It would be very interesting to investigate such a limit using integrability approaches for both correlation functions (hexagonalization \cite{Hex1,Hex2,Basso:2017khq}) and amplitudes (collinear OPE \cite{OPE_for_W_Loops,Basso:2010in_GKP_string,Basso:2013_I,Basso:2013aha_II,Basso:2014koa_III,Basso:2014nra_IV,Basso:2014hfa_All_Helicity_I,Basso:2015rta_All_Helicity_II,Basso:2013vsa_MatrixPart}).

It would also be very interesting to compare expansion (\ref{M5int}) with the results for the five-point correlation function $G_5$ of the 1/2-BPS operators with a large R-charge $K \gg 1$ \cite{Dectagon}. $G_5$ can also be represented as a square of some function which is called the decagon $\mathbb{D}$ if an appropriate projection with respect to the $R$-symmetry indices is chosen: 
\begin{equation}\label{Korr5}
G_5=\frac{\partial^{2K}}{\partial^K\beta\partial^K\gamma}\langle \mathcal{O}(x_1)\ldots \mathcal{O}(x_5) \rangle\Big{|}_{\beta=\gamma=0}= \frac{\mathbb{D}^2}{(x_{12}^2\ldots x_{51}^2)^K},
\end{equation}
where $\mathcal{O}(x_i)=\mbox{Tr}[(y_i\Phi(x_i))^{2K}]$, $y_i$ are auxiliary six dimensional light-like vectors, that parametrise the R-symmetry structure and $\beta,\gamma$ are the coordinates of these vectors in a special basis (see \cite{Dectagon} for details). 
The expansion of $\mathbb{D}$ is known in a special kinematic limit up to two loops \cite{Dectagon}. It contains the $B$, $DB^{(h,v)}$ and $PB$ integrals as well as a new double-box like integral $NDB$ (see fig.\:\ref{fig6} in Appendix \ref{a1}),
which is clearly different from (\ref{M5int}). This is, however, to be expected since a hypothetical duality relation between massive amplitudes and correlation functions will require a small mass (light-cone) limit as well as possible relations between integrals \cite{Eden:2010zz}. In this limit some combinations of integrals may drop out and one may expect to reproduce (\ref{M5int}). In this article, however, we avoid discussing these relations in detail. We hope to address them in a separate publication.

\subsection{Note on a two-point form factor}
\begin{figure}[t]
 \begin{center}
 %\leavevmode
 \epsfxsize=10cm
\epsffile{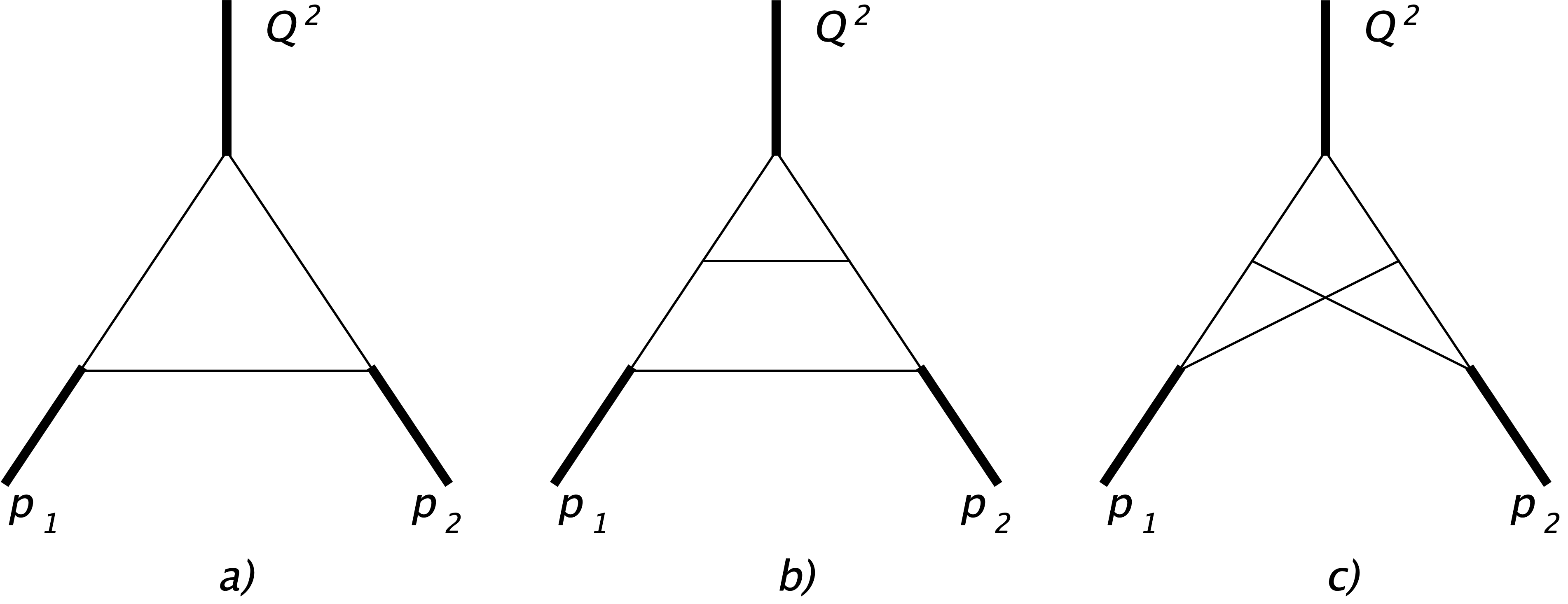}
\end{center}\vspace{-0.2cm}
\caption{Integrals contributing to the stress-tensor supermultiplet form factor at two loops. When external legs are off-shell $a)$ and $b)$ are given by the Ussyukina-Davydychev box functions $(Q^2)^{-1}F_1(p_1^2/Q^2,p_2^2/Q^2)$ and $(Q^4)^{-1}F_2(p_1^2/Q^2,p_2^2/Q^2)$ respectively.}\label{fig51}
\end{figure}
In a pure massless case, the structure of IR poles (\ref{LogMDimRegDiv}) can be related to the behaviour of the two-point form factors of some gauge-invariant operators \cite{Magnea:1990zb,Mueller:1979ih}:
\begin{equation}\label{LogMDimRegDivFF}
\log M_n\Big{|}_{Div.} = \frac{1}{2}\sum_{i=1}^n\log\left( \mathcal{M}_{2}\left(\frac{\mu^2}{s_{ii+1}},a\right) \right)-n\: f(a),
\end{equation}
where in the case of $\mathcal{N}=4$ SYM $\mathcal{M}_2$ can be related to the two-point form factors of the operators from
the stress-tensor supermultiplet \cite{vanNeerven:1985ja,Brandhuber:2010ad,Bork:2010wf}: 
\begin{equation}\label{FF2Div}
\log \mathcal{M}_2\left(\frac{\mu^2}{Q^2},a\right) =-\sum_{l=1}^{\infty}\frac{a^l}{2}\left(\frac{\Gamma^{(l)}_{cusp}}{(l\epsilon)^2}+\frac{G^{(l)}}{(l\epsilon)}\right)\left(\frac{\mu^2}{Q^2}\right)^{l\epsilon} + f(a)+O(\epsilon),
\end{equation}
where $f(a)$ is kinematically independent. Up to two loops this form factor is given by the integrals with the topologies depicted in fig.\:\ref{fig51}. One can also evaluate this form factor in the $\mathcal{N}=4$ SYM on the Coulomb branch with massless external particles \cite{Henn:2011by}. The result for the $\log(m^2)$ terms will be given by (\ref{LogMHigssgDiv}). The topology of the integrals contributing to the form factor remains the same as in the pure massless case and massive particles appear only in loops \cite{Henn:2011by}.

It is interesting to see what happens if we consider massive external states in the small mass limit (from the two-point form factor where on-shell particles are a pair of massive scalars or $W$ bosons). Once again, as in the previous cases, we make a conjecture that the topology of the integrals as well as the values of combinatorial coefficients before them remain the same, external legs are off-shell and internal lines are massless, we see that (here $Q^2$ is the momentum carried by the operator, $Q+p_1+p_2=0$ and $p_i^2=m^2\ll1$, $t=m^2/Q^2$) \cite{vanNeerven:1985ja,Bork:2010wf,Henn:2011by}:
\begin{equation}\label{FF2Div1}
\mathcal{M}_2^{(1)}=F_{1}(t,t),~\mathcal{M}_2^{(2)}=F_{2}(t,t)+\frac{Q^4}{4}NPT(t),
\end{equation}
where $NPT$ is the two-loop Non-Planar Triangle integral depicted in fig.\:\ref{fig51} $c)$. In \cite{Usyukina:1994iw} the following truly magical relation was derived:
\begin{equation}\label{NPTtoF2}
Q^4\:NPT(t)=F_1(t,t)^2,
\end{equation}
which relates this non-planar integral to the square of the one-loop Box integral! Considering $\log \mathcal{M}_2$, we obtain:
\begin{equation}\label{LogF2}
\log \mathcal{M}_2 = -a~F_1+a^2~\left(F_2-\frac{F_1^2}{4}\right)+\ldots
\end{equation}
which after substituting the small mass expansion (\ref{BoxSmallMassExp}) for $F_1$ and $F_2$ results in:
\begin{equation}\label{LogF21}
\log \mathcal{M}_2 = (-a+2\zeta_2a^2+\ldots)\log^2(t)+\tilde{f}(a)+O(m^2),
\end{equation}
where we trace only the $\log^2$ coefficient.
This is exactly
\begin{equation}\label{LogFconject}
\log \mathcal{M}_2 = -\frac{\Gamma_{oct}(a)}{2}\log^2(t)+\tilde{f}(a)+O(m^2)
\end{equation}
expanded up to two loops; $\tilde f(a)$ is some unknown kinematically independent function.
These results are completely in line with the previous discussions, which suggests that the IR structure 
of amplitudes (and form factors) with massive external particles on the Coulomb branch in $\mathcal{N}=4$ SYM is kinematically identical to the cases considered previously in the literature but is governed by $\Gamma_{oct}(a)$ instead of $\Gamma_{cusp}(a)$. Form factors are very interesting objects in $\mathcal{N}=4$ SYM in their own rights and are probably the best objects for
investigation of the origins of the $\Gamma_{cusp}(a) \mapsto \Gamma_{oct}(a)$ rearrangement in the massive case.
Another important, and probably related, question is how the dual conformal invariance manifests itself in this case \cite{Bern:2018oao}. Note that (\ref{NPTtoF2}) relates the non-planar integral with a dual conformally invariant planar one.

\section{Conclusion}\label{s5}
In the current article, we have considered the structure of IR divergences in the case of five-point massive amplitudes in the planar $\mathcal{N}=4$ SYM on the Coulomb branch. We have made a conjecture based on higher-dimensional results \cite{Mafra:2015mja} regarding the structure of the integrand of the five-point amplitude of $W$ bosons in the small mass limit at one and two loops. 

We then have evaluated these integrals at one loop for general kinematics, and at two loops in a single-scale kinematical limit. We have shown that the IR divergences exponentiate and are also controlled by $\Gamma_{oct}$ (at least up to two loops) exactly as in the four-point case considered previously in the literature \cite{Caron-Huot:2021usw}. This allowed us to make a conjecture regarding the all-loop structure of the five-point amplitude which is identical to the four-point case.

It is interesting that if this conjecture is true it will likely imply a hidden simplification in the hexagonalization-based approach to the five-point correlation function \cite{Hex1,Hex2}. 

It would also be interesting to find a connection of our results with a recently proposed generalized Wilson loop description of the amplitudes \cite{Belitsky:2021huz}.

Based on the dual conformal symmetry constraints we expect that at $n>5$ there will be no simple exponentiation pattern for the whole amplitudes in the small mass limit (but IR divergences are expected to exponentiate). This is equivalent to the prediction of the appearance of the finite remainder function in the massless case starting from $n=6$. However, in the light of the results \cite{Basso:2020xts} for the $n=6$ reminder function, we expect major simplifications in the analytical structure of massive (and massless) amplitudes in the "all dual conformal cross-ratios are small" limit for all $n>5$. It would be interesting to investigate this possibility.

\section*{Acknowledgements}
L.V., E.S. and N.B. are very grateful to A.V. Bednyakov for the numerical evaluation
of the integrals with FIESTA.
L.V. is grateful to A.I.Onishchenko and D.I.Kazakov for useful discussions and to 
S.V.Remizov for the help with Linux-based operating systems.  
This work is supported by the Foundation for the
Advancement of Theoretical Physics and Mathematics "BASIS". 
\newpage
\appendix

\section{PentaBox and other integrals}\label{a1}
\begin{figure}[t]
 \begin{center}
 %\leavevmode
  \epsfxsize=5cm
 \epsffile{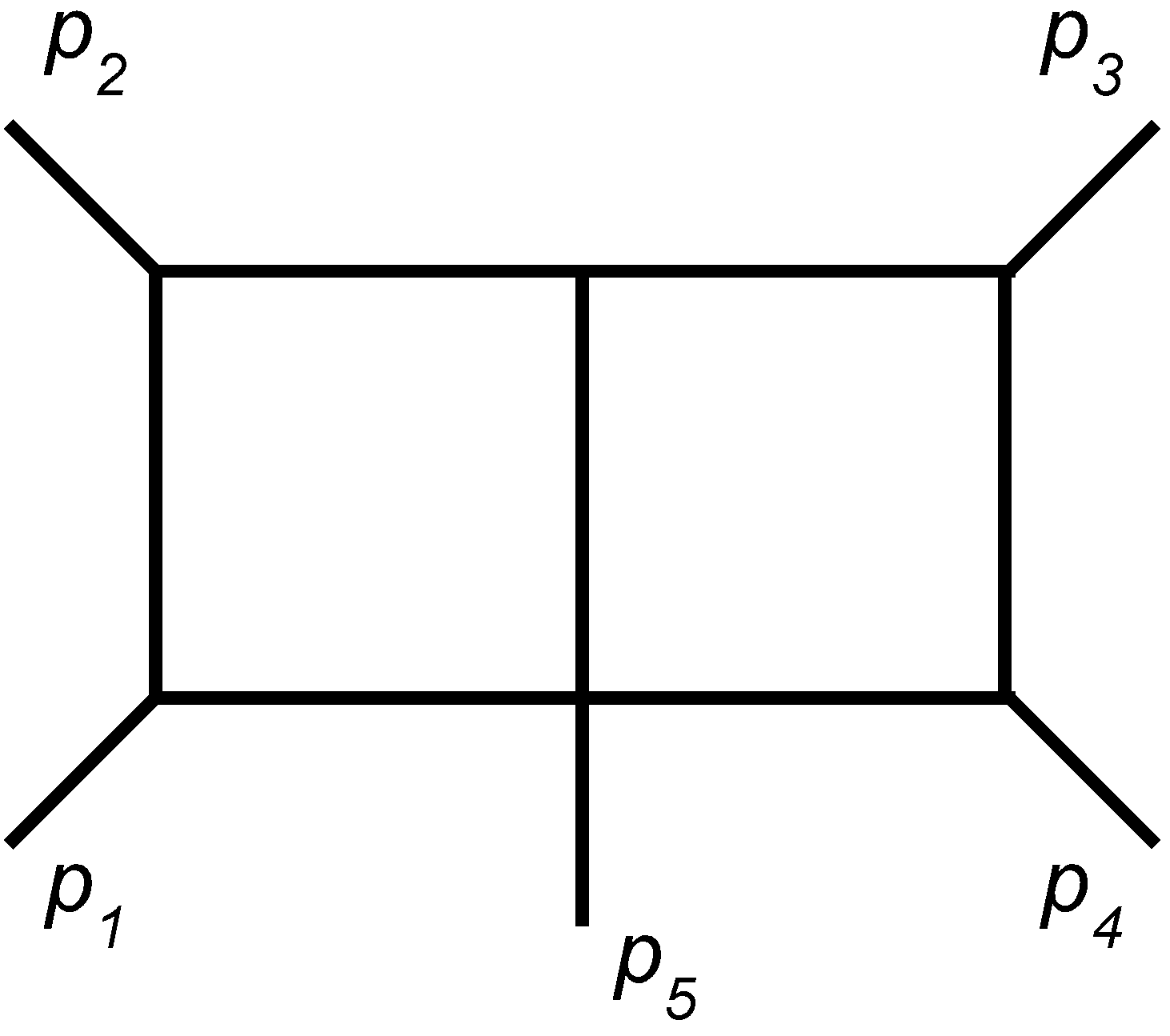}
 \end{center}\vspace{-0.2cm}
 \caption{Five-point "New Double Box" integral $NDB^{(h)}$.}\label{fig6}
 \end{figure}
In the main text we used the Ussyukina-Davydychev box functions in the following form:
\begin{equation}\label{BoxFunct}
\lambda(x,y)=\sqrt{(1-x-y)^2-4xy},~\rho(x,y)=\frac{2}{1-x-y-\lambda(x,y)},
\end{equation}
\begin{equation}
F_l(x,y)=(-1)\sum_{j=l}^{2l}\frac{j!(-1)^{j}\log^{2l-j}\left( \frac{y}{x} \right)}{l!(j-l)!(2l-j)!}
\left( 
\frac{\mbox{Li}_{l}\left(-\frac{1}{\rho~x}\right) - \mbox{Li}_{l}\left( -\rho~y \right)}
{\lambda} 
\right).
\end{equation}
Here $x=p_1^2p_3^2/st$ and $y=p_2^2p_4^2/st$. To consider the equal point $x=y$ case used in the text using (\ref{BoxFunct}), the following (finite) limit must be taken first: $F_l(x,x+\delta)$, $\delta \rightarrow 0$. All other manipulations such as $x\rightarrow 0$ series expansion are straightforward.

For evaluation of the PentaBox integral from the main text let us first define the PentaBox integral with an arbitrary numerator $N$:
\begin{eqnarray}
 PB[N]=\int\frac{d^4ld^4k~~N}{k^2(k+p_1)^2(k+p_1+p_2)^2(k-l-p_5)^2(l+p_5)^2l^2(l-p_4)^2(l-p_4-p_3)^2}.
\end{eqnarray}
The PentaBox integral from (\ref{M5int}) is then given by:
\begin{equation}
PB\equiv PB[(p_1+p_5+l)^2].
\end{equation}
It was convenient for us to split $PB[(p_1+p_5+l)^2]$ into the sum of four integrals:
\begin{equation}\label{PBsum}
PB=DB^{(h)}+PB[2(p_1l)]+PB[2(p_5l)]+s_5~PB[1].
\end{equation}
Then for each of these integrals the Mellin-Barnes (MB) representation was obtained. Throughout all computations of these integrals we used the functionality of MB tools mathematica packages \cite{Gluza:2007rt, Smirnov:2009up, Czakon:2005rk}. See also \url{https://mbtools.hepforge.org}.
For $DB^{(h)}$ the MB representation is 6-fold, while for $PB[2(p_1l)]$, $PB[2(p_5l)]$ and $PB[1]$ integrals the MB representations are 10-fold. For example, the MB representation for $PB[1]$ is (here we use the notation  $z_{i_1 ... i_n} = \sum_{j \in {i_k}} z_j$):
\begin{equation}
\begin{aligned}
PB[1] =&\int \prod_{i=1}^{10} \frac{dz_i}{2 \pi i}\Big(\left(m^{2}\right)^{-4-2 \epsilon-z_1-z_4-z_6-z_8-z_9-z_{10} }(s_{1})^{z_{4}+z_{10}}(s_{2})^{z_9}(s_{3})^{z_1}(s_{4})^{z_6}(s_{5})^{z_8}  \\ %(Q^2)^{z_1+z_4+z_6+z_8+z_9+z_{10}}
&\quad \Gamma(-z_1) \Gamma(-z_2) \Gamma(-z_3) \Gamma(-z_4)\Gamma(-z_5) \Gamma(-z_6) \Gamma(-z_7) \Gamma(-z_8) \Gamma(-z_9) \Gamma(-z_{10}) \\
&\quad  \Gamma(1+z_{1\:3\:5\:8}) \Gamma(1+z_{2\:3\:6\:9}) \\
&\quad \Gamma(3+3 \epsilon+z_{1\:2\:3\:4\:5\:6\:8\:9}) \Gamma(\epsilon-z_7) \Gamma(1+\epsilon+z_{1\:2\:4}-z_7) \\
&\quad \Gamma(-3-2 \epsilon-z_{1\:2\:3\:4\:5\:6\:10}) \Gamma(-2 \epsilon-z_{3\:8\:9}+z_{4\:10}) \\
&\quad \Gamma(1+z_{7\:10}) \Gamma(1-\epsilon+z_{7\:10}) \Gamma(2+z_{4\:5\:6\:7\:10})\Big) / \\
&\quad(\Gamma(-1-4 \epsilon) \Gamma(1+\epsilon-z_7) \Gamma(3+\epsilon+z_{1\:2\:5\:6\:10}+2 z_4) \Gamma(2+z_{7\:10})).
\end{aligned}
\end{equation}
All other MB representations are given in a supplementary Mathematica file. Additionally, any other computational details may be presented if requested. To obtain these representations, we used a 2-fold MB representation of the one-loop Box with three different external masses on external legs and massless internal lines. Then, to simplify computations, we considered a single scale kinematical point $s_i=(p_i+p_{i+1})^2$, $s_i=Q^2$ for $i=1,\ldots,5$, $\mbox{mod}(5)$ and used the functionality of the MB tools packages to generate integration contours and rearrange integrals. Then we used the \texttt{MBasymptotics} package to obtain a small mass expansion. This allowed us to obtain an expansion in powers of $\log(t)$, $t=m^2/Q^2$ 
for each integral. The coefficients before $\log(t)$ were evaluated analytically except for the $PB[(p_1l)]$ case. The constant terms were guessed from the numerical evaluation of MB integrals by means of the \texttt{MBintegrate} function from MB tools. This resulted in:
\begin{equation}
(Q^2)^2~DB^{(h)}\Big{|}_{s_i=Q^2}=\log^4(t)+\frac{13}{2} \zeta_2 \log^2(t)+\frac{21}{2} \zeta_4 +O(m^2),
\end{equation}
which is in agreement with the exact result (\ref{BoxFunct}) for $l=2$. 
\begin{equation}
(Q^2)^3~PB[1]\Big{|}_{s_i=Q^2}=\frac{19}{4}\log^4(t)+16 \: \zeta_2 \log^2(t)-2\zeta_3\log(t)+\frac{83}{4}\zeta_4+O(m^2),
\end{equation}
\begin{equation}
(Q^2)^2~PB[(p_5l)]\Big{|}_{s_i=Q^2}=\frac{5}{8}\log^4(t)+\frac{1}{4}\zeta_2 \log^2(t)-\zeta_3\log(t)-\frac{11}{8} \zeta_4+O(m^2).
\end{equation}
To be on the safe side, we have verified these results for $DB^{(h)}$, $PB[1]$ and $PB[(p_5l)]$ by comparing them with the direct numerical integration of the Feynman parameter integrals using the FIESTA mathematica package
\cite{SmirnovFIESTA} for several kinematical points. The results are in good agreement. 

The integral $PB[(p_5l)]$ satisfies the following identity:
\begin{equation}\label{IntSum}
2\: PB[(p_5l)]\Big{|}_{s_i=Q^2}=NDB^{(h)}\Big{|}_{s_i=Q^2}-DB^{(h)}\Big{|}_{s_i=Q^2}+O(m^2)
\end{equation}
where $NDB^{(h)}$ is represented in fig.\:\ref{fig6}. We have also obtained a MB representation for the
$NDB^{(h)}$ integral, which resulted in:
\begin{equation}
(Q^2)^2~NDB^{(h)}\Big{|}_{s_i=Q^2}=\frac{9}{4}\log^4(t)+7 \: \zeta_2\log^2(t)-2\zeta_3\log(t)+\frac{31}{4} \zeta_4+O(m^2).
\end{equation}
One can see that relation (\ref{IntSum}) indeed holds. The result for $NDB^{(h)}$ was also verified using FIESTA.

For the $PB[(p_1l)]$ integral we found it more convenient to evaluate the coefficients before $\log^2$, $\log$ and the constant term numerically and then fit them with rational numbers and $\zeta_2$, $\zeta_3$, respectively.
We\footnote{We are very grateful to A.V. Bednyakov for the help with FIESTA and for evaluating coefficients before powers of $\log(t)$ in $PB[(lp_1)]$ and $PB$ with appropriate accuracy.} utilize the expansion by regions (see, e.g., Ref \cite{Beneke:1997zp}) and Quasi Monte Carlo integrator \cite{Borowka:2018goh} implemented in the FIESTA5 package 
\cite{Smirnov:2021rhf}. To circumvent problems with spurious singularities in certain regions, we shift all the propagators by an integer multiple of an additional regularization parameter $\lambda$ (see Ref.~\cite{Smirnov:2013eza} for details). %FIESTA computes coefficients of the expansion both in $t$ and $\lambda$.  
After summation of the contributions from different regions, we expand in $\lambda$ to the zeroth order. The obtained results are free from the $\lambda$ dependence and have the form of expansion in $\log t$ with numerical coefficients. With 20 million sampling points we achieve the per mill level relative accuracy\footnote{For the constant terms in the expansion, which are computed with the largest numerical errors.}.
This resulted in
\begin{equation}
(Q^2)^2PB[(p_1l)]\Big{|}_{s_i=Q^2}=-2\log^4(t)- 9 \: \zeta_2 \log^2(t)+2\zeta_3\log(t) - \frac{47}{4}\zeta_4+O(m^2).
\end{equation}
The numerical results from FIESTA are (coefficients that are zero with the current accuracy are omitted):
\begin{eqnarray}
%&&(Q^2)^2~PB[(p_1l)]\Big{|}_{s_i=Q^2}=
(Q^2)^2~PB[(p_1l)]\Big{|}_{s_i=Q^2} &=&  
 -  2. \log ^4(t) - (14.80430\pm 0.00009) \log ^2(t) \nonumber\\
& + & (2.40403\pm 0.00006) \log (t) 
 -  (12.716\pm 0.008)
 + O(m^2).\nonumber\\
\end{eqnarray}

Combining all the integrals in (\ref{PBsum}) together, we finally obtain:
\begin{equation}\label{PBfin}
(Q^2)^3~PB\Big{|}_{s_i=Q^2}=3\log^4(t)+5 \: \zeta_2 \log^2(t)+5 \: \zeta_4+O(m^2),
\end{equation}
which was also independently numerically verified with FIESTA where we have applied the same strategy as for the $PB[(p_1l)]$ case:
\begin{eqnarray}
%&&(Q^2)^3~PB\Big{|}_{s_i=Q^2} =
(Q^2)^3~PB\Big{|}_{s_i=Q^2}  = + 3 . \log ^4(t) + (8.22459\pm 0.00015) \log ^2(t) + (5.412\pm 0.009)
+O(m^2).\nonumber\\
\end{eqnarray}
The results of the direct numerical integration and (\ref{PBfin}) are in good agreement.

\section{Cusp anomalous dimension}\label{a2}
The $\Gamma_{cusp}(a)$ function satisfies the integral BSE equation. The formal solution of this equation can be written as the first element of the inverse power of the product of some infinite-dimensional matrices:
\begin{eqnarray}\label{GCuspSol1}
\Gamma_{cusp}(a)=a\left[\mathbb{Q}\frac{1}{1+\mathbb{K}}\right]_{11},
\end{eqnarray}
where the elements $(\mathbb{K})_{nm}$ and $(\mathbb{Q})_{nm}$ are given by:
\begin{eqnarray}\label{KMatrixElements}
(\mathbb{K})_{nm}=2m(-1)^{m(n+1)}\int^{\infty}_0 \frac{dt}{t}\frac{J_n(\sqrt{2a}t)J_m(\sqrt{2a}t)}{e^t-1},~
(\mathbb{Q})_{nm}=\delta_{nm}\:n\:(-1)^{n+1}.
\end{eqnarray}
In the weak coupling limit, in each order of PT all matrices in (\ref{GCuspSol1}) are effectively finite-dimensional and the coefficients $\Gamma_{cusp}^{(l)}$ can be straightforwardly extracted from (\ref{GCuspSol1}). 

It turned out to be convenient \cite{Basso:2020xts} to partition $\mathbb{K}$ into four blocks by shifting the lines and columns
so that 
\begin{equation}\label{GCuspSol}
\mathbb{K}=
\begin{pmatrix}
\mathbb{K}_{\circ\circ} &  \mathbb{K}_{\circ\bullet} \\
 \mathbb{K}_{\bullet\circ} & \mathbb{K}_{\bullet\bullet} 
\end{pmatrix},
\end{equation}
where the white dot corresponds to the odd index value and a black dot corresponds to the even value.
Then one can consider the deformation of this matrix with the parameter $\alpha$ so that
\begin{eqnarray}\label{GCuspSolMod}
\Gamma(a|\alpha)=a\left[\frac{1}{1+\mathbb{K}(\alpha)}\right]_{11},
\end{eqnarray}
where $\mathbb{K}(\alpha)$ is now given by 
\begin{equation}\label{GCuspSolAlpha}
\mathbb{K}(\alpha)=2\cos(\alpha)
\begin{pmatrix}
\cos(\alpha)\mathbb{K}_{\circ\circ} & \sin(\alpha) \mathbb{K}_{\circ\bullet} \\
\sin(\alpha) \mathbb{K}_{\bullet\circ} & \cos(\alpha)\mathbb{K}_{\bullet\bullet} 
\end{pmatrix}.
\end{equation}
The black and white circle notation is identical to that used previously, and the expression for the $(\mathbb{K})_{nm}$ components is given by (\ref{KMatrixElements}).
Then one can show \cite{Basso:2020xts} that $\Gamma(a|\pi/4)=\Gamma_{cusp}(a)$ and $\Gamma(a|0)=\Gamma_{oct}(a)$. Indeed, expanding (\ref{GCuspSolMod}) one can see that
\begin{equation}
\Gamma(a|\alpha)=2a - 4\cos(\alpha)^2 \zeta_2a^2  + 
 4  \cos(\alpha)^2 (3 + 5 \cos(\alpha)^2) \zeta_4a^3+\ldots,
\end{equation}
which indeed interpolates between (\ref{GammaOctSeries}) and (\ref{GCuspAndGPT}) for $\alpha=0$, $\alpha=\pi/4$ correspondingly.

\newpage
\bibliographystyle{utphys2}
\bibliography{refs_ampl}

\end{document}